\documentclass{article}
\usepackage{arxiv}
\usepackage[utf8]{inputenc} 
\usepackage[T1]{fontenc}    
\usepackage{hyperref}       
\usepackage{url}            
\usepackage{booktabs}       
\usepackage{amsfonts}       
\usepackage{nicefrac}       
\usepackage{microtype}      
\usepackage{lipsum}		
\usepackage{graphicx}
\usepackage{natbib}
\usepackage{doi}
\usepackage{color}

\title{Analysis of core-mantle boundary seismic waves using full-waveform modelling and adjoint methods}

\author {Maria Koroni$^1$\thanks{Corresponding author: maria.koroni@erdw.ethz.ch}, Anselme Borgeaud$^2$,  Andreas Fichtner$^1$ and Fr\'ed\'eric Deschamps$^2$ \\
$^1$ Department of Earth Sciences, Institute of Geophysics, ETH Z\"urich, Sonneggstrasse 5, 8092 Z\"urich, Switzerland \\
$^2$ 
Institute of Earth Sciences, Academia Sinica, 128 Academia Road, Sector 2, Nangang, Taipei 11529, Taiwan
  
}

\date{}

\renewcommand{\headeright}{ }
\renewcommand{\undertitle}{ }
\renewcommand{\shorttitle}{M. Koroni et al., 2021}

\hypersetup{
pdftitle={M. Koroni et al., 2021},
pdfsubject={q-bio.NC, q-bio.QM},
pdfauthor={David S.~Hippocampus, Elias D.~Striatum},
pdfkeywords={First keyword, Second keyword, More},
}

\begin{document}
\maketitle

\begin{abstract}
  Using spectral-element synthetic seismograms and adjoint methods, we present a finite-frequency sensitivity analysis of traveltimes of seismic waves interacting with the core-mantle boundary (CMB). We focus on reflected and refracted \emph{P} and \emph{S} waves that are recorded in seismograms computed using SPECFEM3D\_GLOBE. The core-mantle boundary is the most abrupt internal discontinuity of the Earth, marking the solid-fluid boundary between mantle and outer core that strongly affects the dynamics of the Earth's interior. A lot of research has been dedicated to resolving the structure of the CMB interface, however, a high resolution image of topographic variations is still lacking, due to difficulties relating both to observations in seismograms and good understanding of lowermost mantle velocity structure. We select some of the most important and widely used seismic phases, namely \emph{ScS, SKS, SKKS, PcP, PKP, PKKP} and \emph{PcS}, given their path through mantle and core and their interactions with CMB. These seismic waves have been widely deployed by investigators, trying to comprehend CMB and lowermost mantle structure. Firstly, we use computed synthetic seismograms with a dominant period of \emph{T $\approx$}11s for existing models of CMB topography and compare traveltime perturbations measured by cross-correlation to those predicted using ray theory. We find deviations from a perfect agreement between ray theoretical predictions of time shifts and those measured on synthetics with and without CMB topography. We find that these deviations are generally small when the background velocity model is 1-D, but become larger when a 3-D background model is used, indicating trade-offs between 3-D mantle structure and core-mantle boundary topography. Secondly, we calculate the Fr\'echet kernels of traveltimes with respect to shear and compressional wavespeeds and the boundary sensitivities with respect to the CMB interface. We observe that the overall sensitivity of the traveltimes is mostly due to volumetric velocity structure and that imprints of CMB on traveltimes are less pronounced. In general, higher frequency data is used for imaging CMB, but the intermediate frequency range used here is compatible with current computational capabilities and provides a first approach to an FWI modelling of deep Earth structure. Our study explains the difficulties relating to inferring CMB topography using traveltimes and provides a gallery of exact, finite-frequency sensitivity kernels. We conclude that imaging of CMB using full-waveform inversion (FWI) techniques and kernels as presented here can improve our understanding of deep Earth structure and trade-off between seismic velocities and boundary topography. 

\end{abstract}


\section{Introduction}

The core-mantle boundary (CMB) is a major and sharp discontinuity inside the Earth, separating the mantle, composed of solid rocks, from the outer core, made of liquid iron alloy. Specifically, it lies between the mantle rocky aggregate and the liquid iron alloy outer core. Key phenomena affecting the Earth evolution and dynamics occur on each side of this boundary. For instance, on the mantle side, the CMB region is believed to be the source of mantle plumes, which trigger intraplate volcanism at the Earth's surface. On the core side, magneto-hydrodynamic flow shapes magnetic field lines, partially controlling the magnetic field at the CMB. 
\par
The CMB is the location of interactions between the lower mantle and the outer core \citep[e.g.][]{HeWo01,Stev81,JeanlozWi98,Gurn98,LaWiGa98}. These interactions, in turn, influence mantle and core processes, for instance the geodynamo, and therefore the Earth's magnetic field. Developments of consistent models of the mantle structure and dynamics above the boundary can improve geodynamo modelling, \citep[e.g.][]{BlJa92}. For instance, the heat flux at the CMB, imposed by mantle dynamics, influences the geodynamo, \citep[e.g.][]{Amit2015}. Lateral variations of the core-mantle boundary radius, with its aspherical structure, may affect outer core flow processes, implying that a better image of CMB topography is essential to analyses of gravitational and topographical coupling between mantle and core, \citep[e.g.][]{GuRi86,Calkins2012,DaStDu14}.
\par
From a seismological perspective, the core-mantle boundary is one of the most interesting regions for interdisciplinary geophysical research, however it is difficult to resolve it robustly. Over the years, much seismological research has been produced regarding lower-mantle and core-mantle boundary topography structure, using body waves and secondary phases of \emph{P} and \emph{S} waves, \citep[e.g.][]{Creager1986,Doornbos1989,Shen2016,Schlaphorst2016,Wu2014,Restivo2006,Sze2003,Morelli1987,ObFu97,RoWa93,Mancinelli2016,Tanaka2010} as well as normal modes, \citep[e.g.][]{IsTr99,Xiang-DongLi1991,Soldati2013}.
However, existing discrepancies among suggested CMB topography models for spherical harmonic degrees higher than two \citep{BeBo02,Koe20} indicate that there may be features of the lowermost mantle structure, which need to be taken into account systematically. Additionally, possible inaccuracies in interpreting observed data can also be a reason for these differences, as most studies usually do not jointly explain CMB topography and velocity structure. 
\par
An important hindering factor is the complexity of the seismic structure in the bottom 300-400 km of the mantle. This includes the D" discontinuity, which is not present everywhere and is thought to be related to the phase transition from bridgmanite to post-perovskite, \citep[e.g.][]{Hirose2017,Cobden2015}, ultra-low seismic velocity zones (ULVZ), which are thin pockets where compressional and shear wave velocities ($V_{P}$ and $V_{S}$) drop by up to 10 \% and 30 \%, respectively, \citep[e.g.][]{YuGa2018}, strong, small-scale anomalies induced by cold oceanic slabs that have sunk down to the CMB, \citep[e.g.][]{Hung2005,Whittaker2015,Borgeaud2017}, and the large low shear-wave velocity provinces (LLSVPs), defined as regions a few thousands of kilometers across where shear velocity drops by a few percent, and which are located beneath Africa and the Pacific ocean, \citep[e.g.][]{LiRo96,s40rts,LeCoDzRo12,GaMaSh16,Mac18}. Because they may be closely related to mantle dynamics, LLSVPs may strongly influence CMB topography. Numerical simulations showed that depending on their exact nature, clusters of purely thermal plumes or thermo-chemical piles, LLSVPs would trigger either uplifts or depressions in the CMB \citep[]{Lassak2010,Deschamps2018}.  
\par
Ideally, one would like to resolve CMB topography structure, while simultaneously obtaining an updated velocity model for the mantle. It has been shown by many researchers that CMB topography and mantle velocity variations exhibit strong trade-off, which makes the inference of a topographic map along the core-mantle boundary a hard task, \citep[e.g.][]{GaSo00,Sze2003}. 
\par 
The nature of difficulties in resolving lowermost mantle structure can be analysed and addressed by investigating seismic waves which are sensitive to lowermost mantle structure and have been used to constrain the topography of the CMB. These seismic phases are characterised by complex wave phenomena occurring while travelling from the source to receivers, which makes it cumbersome to determine which type of seismic waves has a maximum CMB sensitivity. 
Nonetheless, there have been many studies analysing the compressional and shear wavespeed structure of the lowermost mantle using several related body wave phases as well as differential traveltimes between these, \citep[e.g.][]{LiRo96,Gar00,GaMaSh16,s40rts,Tanaka2002,LeCoDzRo12,MuTk20}. A summary of the progress made during the last decades and relevant literature can be found found in \citep{Gurn98,Gar00} and \citep{Koe20}.
In a different approach, studies \citep{Colombi2012,Colombi2014}, based on axisymmetric spectral-element modelling \citep{axisem}, have implemented boundary and volume sensitivity kernels within a spectral-element code given spherically symmetric background models. They used this implementation to address the sensitivity of \emph{PcP} and \emph{P$_{diff}$} and showed a suite of sensitivity kernels to CMB structure \citep{Colombi2012}. Their research gave insights into how to incorporate sensitivity of traveltimes of \emph{P$_{diff}$, PKP, PcP} and \emph{ScS} for imaging the CMB \citep{Colombi2014} in a joint inversion for boundary and lowermost mantle velocity structure in order to tackle the trade-off between parameters. Our study is stepping further by using the full-waveform synthetics calculated with spectral-elements in a cubed sphere, as implemented in SPECFEM3D\_GLOBE, where the global wave propagation effects can be more accurately accounted for and implemented into the full-waveform processes in a consistent way.
\par
The goal of this work is to investigate the finite-frequency sensitivity of frequently used seismic phases for lowermost and core-mantle boundary structure and provide a kernel gallery showing their sensitivity to boundary and volumetric isotropic structure. This is following the formulation of finite-frequency traveltime sensitivity given by \citep{marquering1999,Dah05} and \citep{TrTaLi05} with the use of adjoint methods. The main contribution of this work is to better understand how to optimally use traveltime data of particular \emph{P-} and \emph{S}-waves for imaging the CMB and show how their sensitivities can be incorporated in a full-waveform inversion scheme that is targeted for resolving the lowermost mantle and CMB topography. We investigate a suite of \emph{P-} and \emph{S}-wave phases and their traveltimes as observed in synthetic seismograms computed for intermediate frequencies (0.01-0.08 Hz), which are at the higher end of frequencies used in global waveform inversion studies. We focus on the seismic phases \emph{PcP, PKP, PKKP, PcS, ScS, SKS} and \emph{SKKS}. We deploy a full-waveform, finite-frequency approach for analysing CMB related seismic waves, using well-established techniques based on spectral-element waveform modelling and adjoint methods for calculating the Fr\'echet sensitivity kernels, \citep[e.g.][]{KoTr99,KoTr02a,KoTr02b,TrTaLi05}. 
\par
To address the effects of adding topography and/or 3-D velocity variations on the traveltimes of these phases, we compare existing ray-based methods of inferring core-mantle boundary topography to cross-correlation time shift measurements made on the spectral-element synthetics. For this, we use full-waveform synthetic seismograms for events selected especially for CMB topography studies. The analysis is made by comparing traveltime anomalies predicted using ray theory, for the CMB topography models mentioned above, to residual traveltimes measured on full-waveform synthetics computed for the same CMB topography models. The traveltime comparisons are done similar to \citep{BaZhRi12} and \citep{KoTr16}. This comparison could provide insights into the usability of the traveltimes of our selected phases within the ray theoretical framework. Using full-waveform modelling as realistic recordings, we can address the effects of varying the model on specific observables. 
This analysis should benefit from investigating the separate effects of 3-D mantle velocity structure and boundary topography on the traveltimes, when only these model parameters are perturbed during the simulations. The separate effects are examined by computing synthetics in a 3-D model and in a 1-D velocity model plus core-mantle topography.
Our results show that for 1-D background the ray-theoretical framework works better for \emph{S} phases than for \emph{P} phases, while it deviates from a good prediction when the background velocity model is 3-D. We also find that the boundary sensitivity of almost all phases is much less pronounced than their clearer sensitivity to volumetric parameters, both compressional and shear velocity. This confirms the reported trade-off between CMB and 3-D velocity structure, and also indicates that all types of kernels shown here could be used to provide better constraints on relevant structures, as the CMB topography cannot solely be responsible for the traveltime shifts and the velocity structure imprint is significant. 
\par
The paper is structured as follows: We first present the methods for carrying out the traveltime analyses and comparisons. Then we explain the specifics of Fr\'echet sensitivity analysis. We present our results and we discuss them by addressing the effects of different CMB topography models, the influence of the background velocity model (1-D or 3-D). We also consider the different effects on \emph{P-} and \emph{S-} seismic phases. Finally, we explain the resulting Fr\'echet kernels in the context of full-waveform inversion and trade-off between the two parameters.

\section{Methodology}
\subsection{Computation of synthetic seismograms}

\begin{figure}
    \centering
    \includegraphics[width=.75\textwidth, angle=90]{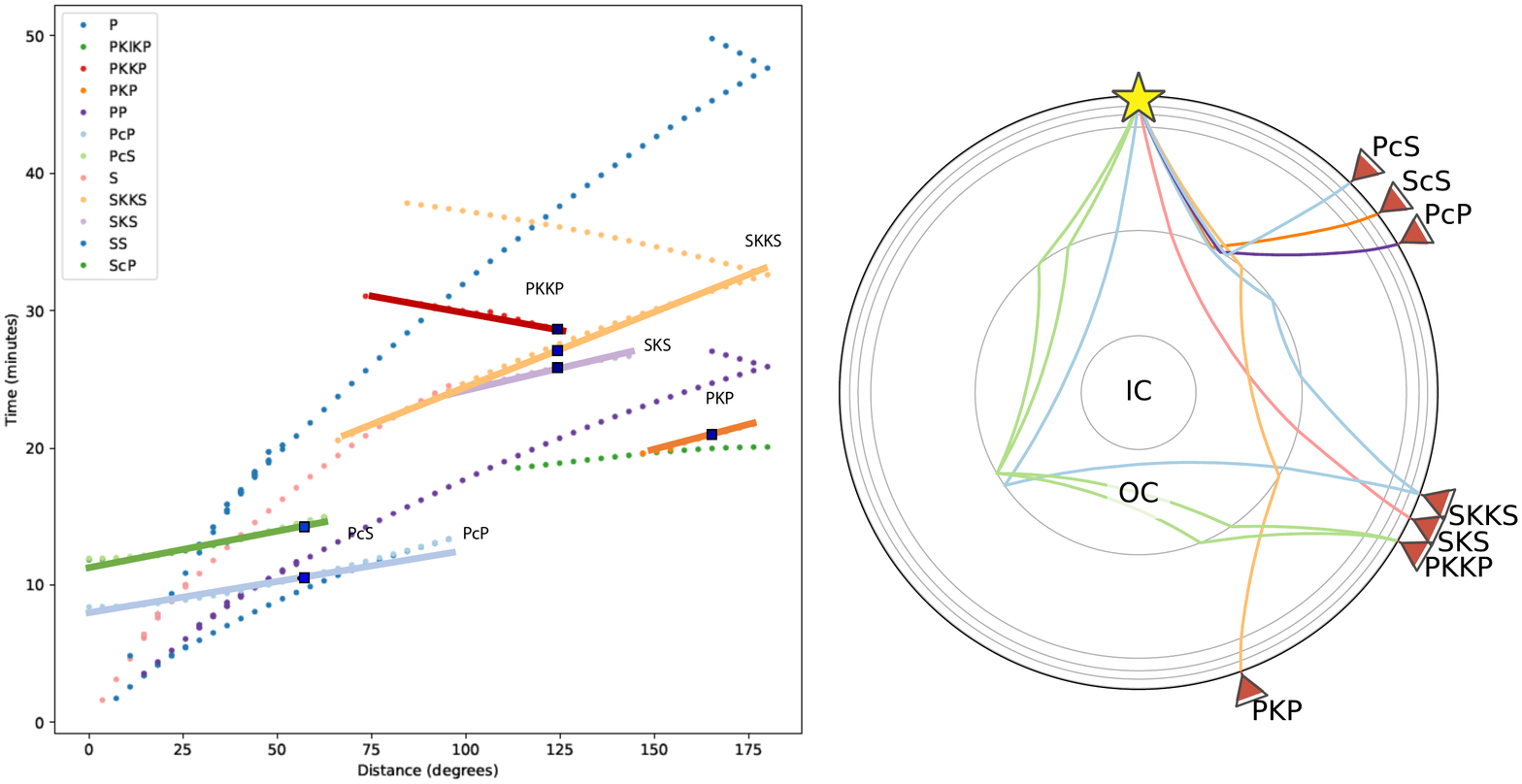}
    \caption{\textit{Left}: Traveltime curves according to model PREM and tauP \citep{tauP}. The phases we investigate in this work are drawn with solid lines. More primary phases are plotted in order to justify the chosen epicentral distances for the sensitivity analysis. The characteristic distances chosen for computing the sensitivity kernels are shown by blue circles on the solid lines for each phase. \textit{Right}: Theoretical raypaths for the same model. Red triangles denote stations at the epicentral distances were we opt observing the phases for the sensitivity analysis. The colours in the two sub-figures are not corresponding. IC denotes inner core, OC denotes outer core.}
    \label{fig:raypaths}
\end{figure}

In order to enhance our knowledge about core-mantle boundary structure by investigating its effects on \emph{P-} and \emph{S-} seismic phases, we perform full-waveform modelling and focus on frequencies ranging from 0.01-0.08 Hz, given the computational limitations, still lying within an acceptable observational resolution. 
The theoretical raypaths of the chosen seismic phases are shown in Figure $\ref{fig:raypaths}$, which is produced using 1-D ray tracing with tauP \citep{tauP} and using model PREM \citep{DzAn81} as reference. 
The computation of synthetic waveforms is done using the community code 
SPECFEM3D\_GLOBE \citep{KoTr02a,KoTr02b} in 1-D and 3-D mantle velocity models, namely PREM \citep{DzAn81} and S20RTS \citep{RiVh02}, respectively. The synthetic waveforms are used to imitate realistic waveforms for the comparisons between ray theory and measurements and to assess the direct effects of CMB topography and 3-D velocity variations on the traveltime of the seismic phases.
\par
\begin{figure}
    \centering
    \includegraphics[width=\textwidth]{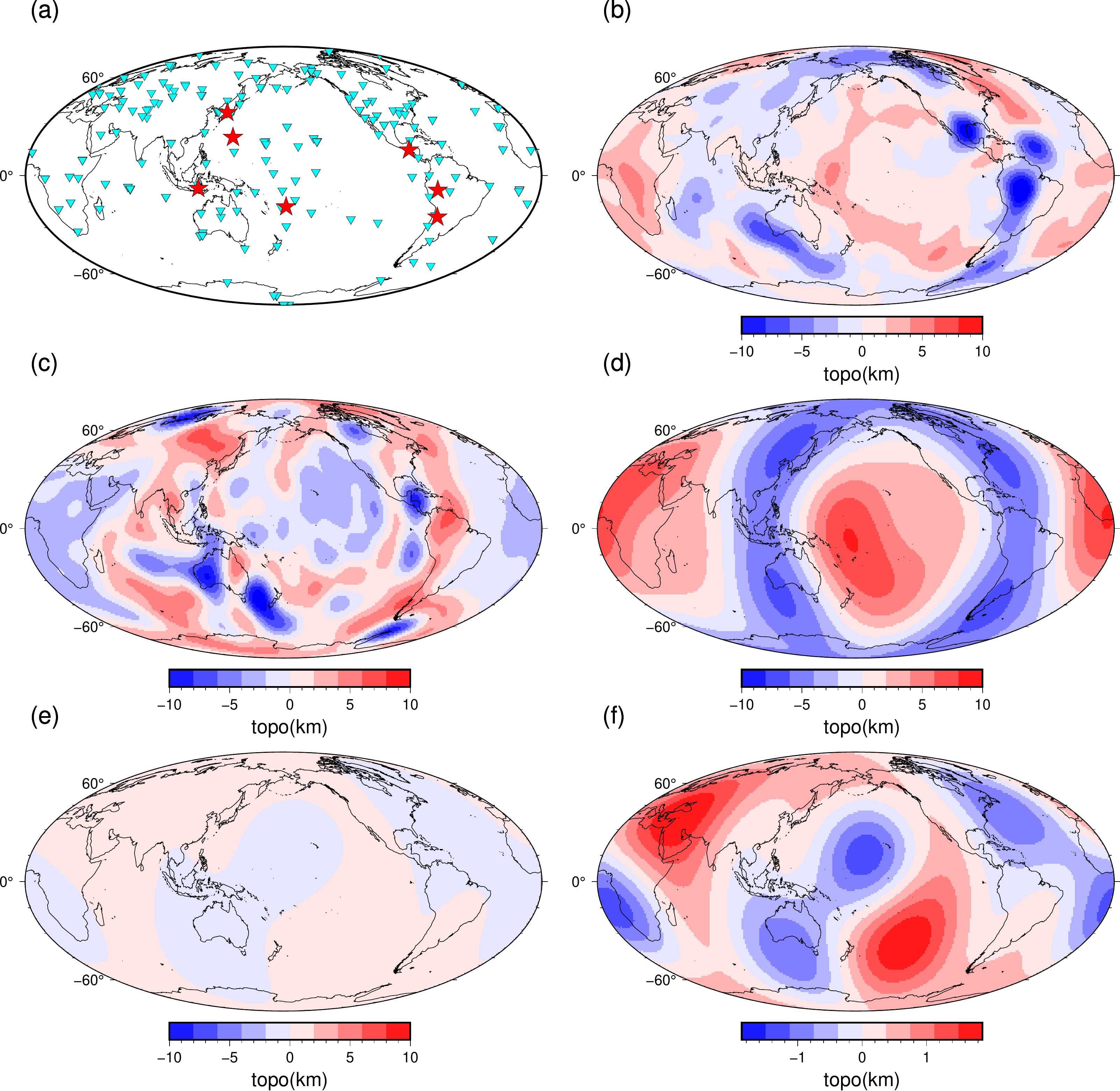}
    \caption{Maps of the earthquake events and stations used in this study (upper left). Maps of the CMB topography models used in this study are shown as well. 
    Specifically, panel (a) shows the events (red stars) and stations from the Global Seismographic Network (cyan inverted triangles). Panels (b) to (e) show models \emph{{T1}} (b) and \emph{{TC1}} (c) based on the mantle convection simulations of \citep{Deschamps2018} (see main text), and \emph{{LM91}} \citep{Xiang-DongLi1991} (d) and \emph{{TK10}} \citep{Tanaka2010} (e), inferred from actual data. Panel (f) shows the same model \emph{{TK10}} as in panel (e), except for this time there is a  change in scale of the colour bar, as this model's topographic variations are much smaller than the other models.}
    \label{fig:models}
\end{figure}

For the comparisons of traveltimes derived using ray theoretical methods to measurements done on full-waveform synthetics, existing core-mantle boundary topography models are added, resulting in PREM or S20RTS plus CMB topography. A model with very low topographic variations is used, derived by \citep{Tanaka2010} referred here as \emph{{TK}}. In order to examine the effects of a variety of topographies with larger perturbations, we compute synthetic waveforms by also adding the models by \citep{Xiang-DongLi1991}, denoted here as \emph{{LM91}}, with peak-to-peak variation scaled up to 8 km, and two topography models derived from geodynamic simulations \emph{{T1-pPv}} and \emph{{TC1-pPv}} from \citep{Deschamps2018}, denoted here as \emph{{T1}} and \emph{{TC1}} for simplicity. In order to make the geodynamics models more applicable to our source-receiver geometry, model TC1 is rotated so that its spherical harmonic degree-2 pattern match best that of S20RTS S-velocity variations right above the CMB (this is done by performing a simple grid search). The same is done for model T1, but with the sign of S-velocity variations inverted. This is because S-velocity anomalies at the CMB, and CMB topography variations are expected to be positively correlated for thermo-chemical models (TC1), but negatively correlated for purely thermal models (T1) \citep{Deschamps2018}.
These topography models are derived from purely thermal (\emph{{T1}}) and thermochemical (\emph{{TC1}}) simulations of mantle convection modelling. They were filtered for spherical harmonic degrees $\ell$ = 0 to $\ell$ = 20, and have absolute peak-to-peak amplitude around 20 km, for both models. 
Most of this amplitude is however accommodated by deep depressions caused by downwellings reaching the CMB, while positive topography associated with plume clusters in \emph{{T1}} and depressions associated with thermochemical piles in \emph{{TC1}} have height or depth, respectively, of about 2 km. Having various possible topography models allows us to investigate the reliability of ray theoretical methods in a more conclusive manner. All CMB topography models used here along with the dataset constructed by the earthquake events shown in table $\ref{tab:events}$ are displayed in Figure $\ref{fig:models}$. 
\par
For the simulations, we set the resolution such that the resulting noise-free waveforms have dominant period of about $T={11}$ \emph{s}. Some modelling parameters which are present in real data can be important, such as attenuation, and can have complex effects on traveltimes of seismic phases, thus they are not considered for our simulations, since we want to focus on the effects of CMB topography and 3-D lateral isotropic variations without further complicating the waveforms. 
\par
To obtain a realistic dataset, we use seven events with a range of focal depths, source time function duration and moment magnitudes in the range $Mw=5.5-7.5$. The events are summarised in table $\ref{tab:events}$. The receiver locations are set to those of the Global Seismographic Network and we make measurements for each source-receiver pair as well as ray theoretical predictions for the same pairs. This is done in time windows around the seismic phases of interest. The synthetic data are processed by filtering using a band-pass between the frequencies 0.01-0.08 Hz. The synthetic waveforms are used for two different purposes: Firstly, for comparisons of traveltime shifts caused by topographic variations in 1-D or 3-D mantle structure models and, secondly, for computing the sensitivity kernels for selected time windows encapsulating the maximum peak of the energy of our selected seismic phases. 

\begin{table}
\footnotesize
    \begin{tabular}{l|l|l|l|l|l|l}
    	 \textbf{GCMT ID} & \textbf{Event Location} & \textbf{M$_w$} & \textbf{Latitude ($^\circ$)}  & \textbf{Longitude  ($^\circ$)} & \textbf{Depth (km)} & \textbf{STF duration (\emph{s})} \\ [.1ex] \hline \hline
         200609090413A & Flores Sea & 6.3 & -7.23 & 120.27 & 583.2 & 6.8  \\ 
         200707161417A & Sea of Japan & 6.8 & 36.84 & 135.03 & 374.9 & 12.4 \\
         200709281338A & Volcano Islands (JP) & 7.4 & 21.94 & 143.07 & 275.8 &  26.4 \\
         201409241116A & Jujuy Province (AR) & 6.2 & -23.78 & -66.72 & 227.6 & 6.4 \\
         201303252302A & Guatemala & 6.2 & 14.62 & -90.71 & 186.4 & 6.2 \\ 
        200602021248A & Fiji Islands & 5.9 & -17.7 & -178.13 & 611.6 & 11 \\
         101202H & Western Brazil & 6.9 & -8.30 & -71.66 & 539.4 & 13.2 \\
    \end{tabular}
    \caption{Main features of the earthquake events used in this study. These are recorded at the station locations of the Global Seismographic Network.}
    \label{tab:events}
\end{table}{}

\subsection{Comparison of ray theoretical time shift prediction and cross-correlation measurements}\label{sse:rt_vs_fwt}

In this section, we describe the prediction of traveltime anomalies due to perturbations of CMB topography using ray theory (RT), and as measured on full-waveform synthetics computed using the spectral element method (denoted by FW, which stands for full-waveform) using cross-correlation. The motivation of this analysis is to check whether we can reliably predict a topographic perturbation given the ray-theoretical framework. If this is true, the measured by cross-correlation time shift and the predicted time anomaly for a specific topographic variation should be identical. In the opposite case, where the two values diverge, there are two implications: firstly, the ray-theoretical framework is not sufficient for translating the time shift measured to topographic variation; secondly, there are velocity effects which are not adequately taken into account when using a linearised inversion. Traveltime perturbations due to the CMB topography are computed in the framework of ray theory following \citep{Morelli1987,Tanaka2010}, using the expressions given in eq. $\ref{eq:rt_kernel_ref}$ when the ray reflects off the CMB (e.g., \emph{ScS}), and eq.  $\ref{eq:rt_kernel_tra}$ when the ray transmits through the CMB (e.g., \emph{SKS}):

\begin{equation}\label{eq:rt_kernel_ref}
\delta t_{ref} = -\frac{2}{r}\left( \eta^2 - p^2 \right)^{\frac{1}{2}} \delta r,
\end{equation}

\begin{equation} \label{eq:rt_kernel_tra}
\delta t_{tra} = -\frac{1}{r} \left[ \left( \eta_{+}^2 - p^2 \right)^{\frac{1}{2}}
- \left( \eta_{-}^2 - p^2 \right)^{\frac{1}{2}} \right] \delta r,
\end{equation}

where $\delta r$ (in \emph{km}) denotes the variation of the CMB radius at the bouncing points, $p$ (in \emph{s/rad}) is the ray parameter for the 1-D model without perturbations of CMB topography, and $\eta = r_{cmb} / v(r_{CMB})$ (in \emph{s/rad}) is the vertical ray parameter, with $\eta_{+}$, and $\eta_{-}$ representing the values of $\eta$ just above, and below the CMB, respectively. 
For PREM, which is our selected model, $\eta_{+}=479.03$ for \emph{S} phases, and $\eta_{+}=253.71$ for \emph{P} phases, while $\eta_{-}=431.51$ (only \emph{P} phases are supported in the core).
\par
Traveltime perturbations for the synthetics are computed by cross-correlation between synthetics for the reference model (either PREM, or S20RTS) and for the same model with added perturbations of CMB topography. 
For each of the seismic phases \emph{ScS, SKS, SKKS, PcP, PKP, PKKP} and \emph{PcS}, we first compute preliminary time windows from 5~s before, to 20~s after the arrival time of the phase as predicted using tauP \citep{tauP} and PREM. We exclude time windows that overlap with the phases \emph{S, SS, SSS, sS, P, pP, PP, PPP, sP} and \emph{pS}, since overlapping phases affect the traveltime measurement by cross-correlation, which makes the comparison with ray theory less meaningful.
\par
\begin{figure}
    \centering
    \includegraphics[width=.75\textwidth, angle=0]{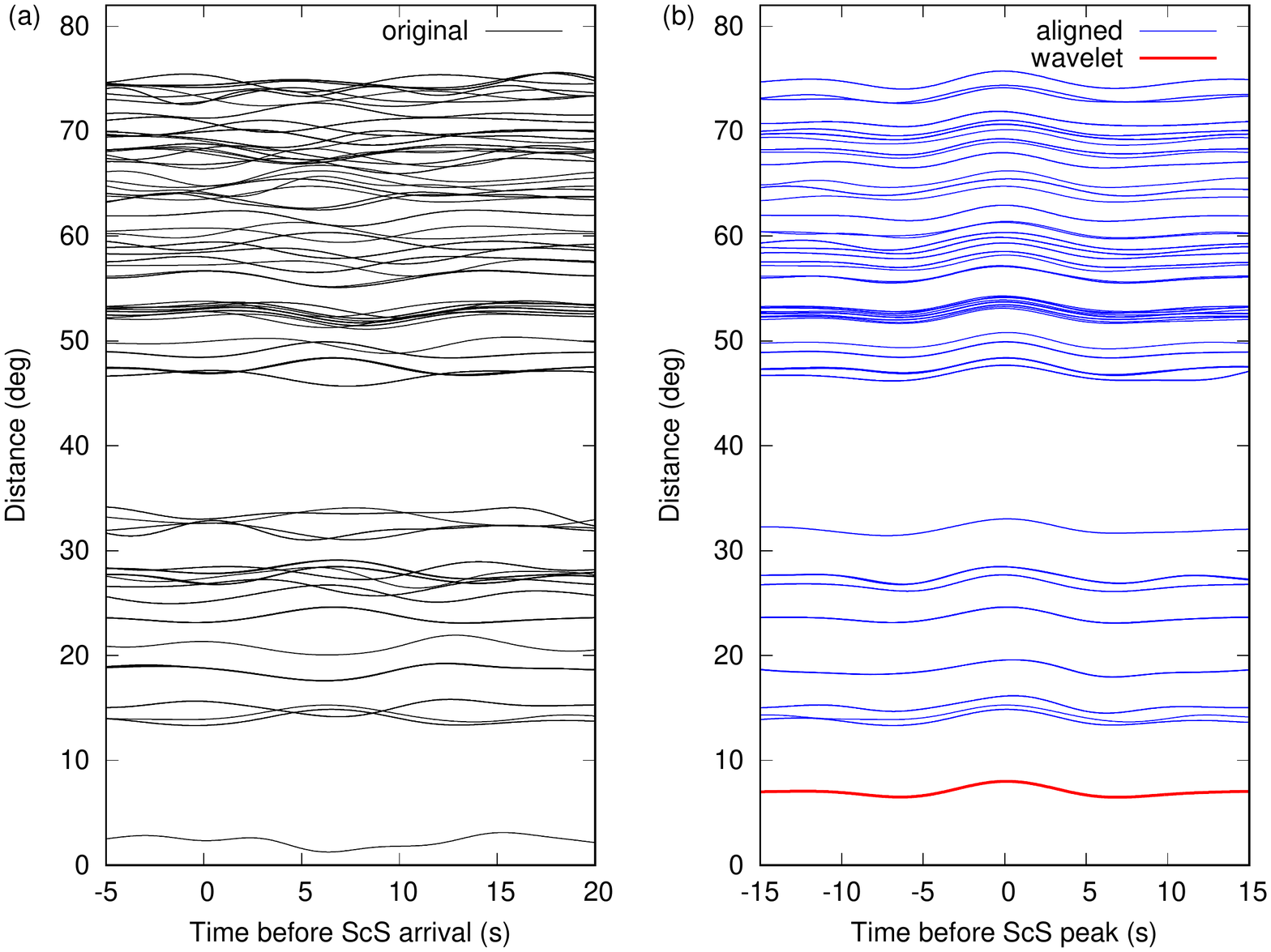}
    \caption{Illustration of the adaptive stacking procedure \citep{Rawlinson2004} used to define the time windows, for the ScS phase for model S20RTS and event no. 1 in Table $\ref{tab:events}$. (a) Waveforms before alignment, using time windows 5~s before, and 20~s after the ScS arrival; (b) same as (a), but after alignment using the source wavelet (red trace) obtained from the adaptive stacking procedure.}
    \label{fig:iterstack}
\end{figure}

For a given reference model, we then compute improved time windows centered on the peak of each phase following the adaptive stacking algorithm of \citep{Rawlinson2004}, in order to: 1) correct for the shift in travel-time for the case of a 3-D reference model, and 2) further eliminate time windows with overlapping phases. 
The improved time windows have a length that ranges from -15 s to +15 s after the arrival time of the peak for each phase. The output of the adaptive stacking algorithm is the time windows aligned on their respective peak amplitude and an average source wavelet that represents the apparent source time function of the phase of interest. An example of the adaptive stacking procedure is shown in Figure \ref{fig:iterstack}.
The selection of time windows without overlapping phases is done by imposing a minimum cross-correlation coefficient between a given time window, and the inferred source wavelet. In the remaining of this section, the minimum cross-correlation coefficient is set to 0.8 for the \emph{P} phases (\emph{PcP, PKP, PKKP} and \emph{PcS}), and to 0.95 for the \emph{S} phases (\emph{ScS, SKS, SKKS}), due to the higher signal-to-noise ratio for \emph{S}- than for \emph{P}-waves. For the \emph{PcP} phase, we further exclude all records at epicentral distances smaller than 40$^{\circ}$, since such records show consistent, large disagreement between RT and FW. The total number of time windows before and after selection is shown in Table \ref{tab:num_windows}. The amount of time windows that pass our quality criteria becomes significantly smaller, however, we prefer to base our conclusion on a more robust dataset. The smaller numbers of high correlation coefficients time windows in noise-free synthetics also shows that the isolation of these phases can be considerably cumbersome in real datasets.
\begin{table}
    \centering
    \begin{tabular}{c |c| c| c}
        \hline
        Phase & Before selection & Selected (PREM) & Selected (S20RTS)\\
        \hline
        \hline
        ScS & 1273 & 795 & 550\\
        SKS & 2262 & 932 & 905\\
        SKKS & 2296 & 587 & 666\\
        PcP & 570 & 137 & 151\\
        PKP & 375 & 322 & 264\\
        PKKP & 1715 & 700 & 775\\
        PcS & 889 & 449 & 379\\
        \hline
    \end{tabular}
    \caption{Number of time windows used in this study, for each phase before selection based on cross-correlation coefficients, and after selection for the PREM and S20RTS models. Note the smaller number of PcP time windows before selection compared to ScS, which is in part due to \emph{P} merging with \emph{PcP} at smaller epicentral distances than for \emph{S} and \emph{ScS}.}
    \label{tab:num_windows}
\end{table}

\subsection{Computation of Fr\'echet sensitivity kernels}

The sensitivity kernels are computed for time windows of 30 \emph{s} length, centered on the predicted peak arrival of the \emph{P-} and \emph{S}-wave phases under investigation. The sensitivity is computed based on the Born approximation \citep[e.g.][]{marquering1999,DaHuNo00,Dah05} and using adjoint methods as implemented in the software package SPECFEM3D\_GLOBE for boundary \citep{Dah05,Liu2008} and volumetric kernels \citep{TrTaLi05}. With this analysis, we conduct an investigation of the traveltimes by visualising their sensitivity to volumetric, namely shear and compressional velocity mantle structure, and boundary parameters, this is interface denoting CMB. The main goal is to distinguish these phases which can truly help us to improve core-mantle boundary topography models in a full-waveform inversion scheme, judging by the effect this has on their traveltime. 
\par
More specifically, the kernels are computed with respect to the CMB radius and compressional and shear wavespeed in 1-D background model PREM.
We compute the so-called "banana-doughnut" traveltime kernels , \citep[e.g.][]{LuSc91,marquering1999,DaHuNo00}, which associate a traveltime measurement on the waveforms to the changes in the finite-frequency sensitivity depending on the background model. 
The traveltime refers to the time window around the arrival of the phase according to a given model, which we have fixed to PREM, for a single pair of source-receiver. 
The relationship connecting a traveltime window to volumetric structure defined by seismic properties of compressional ($\alpha$) and shear ($\beta$) wavespeed are shown below. The following expressions are given for a reference seismic receiver ($r$) at position $x_r$. The reader is referred to \citep{TrTaLi05} for the complete derivations according to the Born approximation and the adjoint method. The volumetric kernels are related to the time window via:
\begin{equation}
\label{eq:vol}
\delta T_{r} = \int _{V} K_{\alpha, \beta}(\mathbf{x},\mathbf{x_{r}}) \delta \ln \alpha , \beta(\mathbf{x}) d^{3}\mathbf{x},
\end{equation}

The units of volumetric wavespeed kernels are $s \cdot km^{-3}$.
\par
The expression relating the traveltime window which encloses the seismic phase to the boundary perturbation on the core-mantle boundary (solid-fluid interface denoted by $\Sigma F$) is based on theory derived by \citep{Dah05} and implemented in the spectral-element method:
\begin{equation}
\label{eq:bound}
\delta T_{r} = \int _{\Sigma F} \mathbf{K}_{h} \cdot \delta h(\mathbf{x})d^{2}\mathbf{x},
\end{equation}
whereby $\mathbf{{K}_{h}}$ is the time integration for variations along the boundary \citep{Dah05}. The units of boundary kernels are $s \cdot km^{-2}$.
\par
We select the phases for an ideal set-up of source-receiver given their epicentral distance also according to literature and optimal theoretical observational range. Specifically, for phases \emph{ScS, PcP}, and \emph{PcS} the epicentral distance is chosen to be 53$^{\circ}$, as at this distance they are less likely to interact with \emph{SS, PP}, and \emph{P}, respectively. The phases \emph{PKKP, SKS}, and \emph{SKKS} are observed at an epicentral distance equal to 110$^{\circ}$ for better isolation of their theoretical arrival time. 
The \emph{PKP} phase is selected at a distance of 160$^{\circ}$ as, according to 1-D ray theory and tauP analysis, at this distance, it is better separated from \emph{PKIKP}. The phases that travel through the outer core as well as \emph{PcS} are time windowed on radial components of the synthetic seismograms, as they are expected to have maximum energy at this component, while core reflected phases are observed and analysed on the vertical (for \emph{PcP}) and transverse component (for \emph{ScS}). 
The traveltime curves are shown in figure \ref{fig:raypaths} with annotated characteristic distances for each seismic phase window. 
\par
For the visualisation of the sensitivity kernels, we use a characteristic scale length, which is appropriate for the range of data values of the kernels and allows us to observe the theoretical finite-frequency sensitivity. 
It is helpful to consider that the volumetric and boundary kernels have different units and are explicitly calculated for different parameters; therefore the model update using these kernels will be different in the context of full-waveform inversion, depending also on the optimisation algorithm chosen for the potential model updates. Considering the expression given above, i.e. eq. $\ref{eq:vol}$ and eq. $\ref{eq:bound}$, it is not easy to directly compare boundary to volumetric kernels
and assess how they will change the initial model. It is however meaningful to consider that the traveltimes of the selected seismic phases will likely show sensitivity to both parameters, as expected by the potential trade-off between velocity and CMB structure, and thus it is worthwhile to judge the qualitative effects of volumetric and boundary structure separately.

\section{TRAVELTIME ANALYSIS USING FULL-WAVEFORM MODELLING}
\label{sec:dts}
 In this section, we show the comparisons of traveltime anomalies measured on FW synthetics to RT prediction for phases \emph{ScS, SKS, SKKS, PcP, PKP, PKKP}, and \emph{PcS}. In Figures $\ref{fig:Pwaves-1D-models-RTWF}$ and $\ref{fig:Swaves-1D-models-RTWF}$, we present the results of this comparison for all topography models and for all the selected phases, for synthetic waveforms calculated using PREM, and S20RTS (1-D, and 3-D, respectively) as velocity background model. Figures $\ref{fig:Pwaves-1D-models-RTWF}$ and $\ref{fig:Swaves-1D-models-RTWF}$ show that there is part of the measured traveltime shift which is not fully recovered when translating a topographic variation to time delay using ray theory. 
 This essentially means that the ray theoretical expression cannot fully explain the topographic variation as time anomaly measured in full-waveforms with and without topography perturbation. The discrepancy seems to increase linearly with $dt_{RT}$, and thus with the amplitude of the CMB topography, and varies significantly for different topographic models and phases. If the recovery using RT could accurately explain the traveltime shifts on these seismic phases, all points in Figures $\ref{fig:Pwaves-1D-models-RTWF}$ and $\ref{fig:Swaves-1D-models-RTWF}$ would lie on the line with slope equal to one. However, the slopes of the linear regression lines have values different from one, as shown in Table \ref{tab:dt_regression}. For the \emph{ScS} phase (which gives the most corresponding results between RT predictions and WF measurements) with PREM as a background model, deviations from a perfect agreement are between 2\% to 7\%, while other well-resolved phases (\emph{SKS, SKKS, PcP} and \emph{PKP}) show deviations of up to 26\% (for the \emph{SKS} phase for model \emph{T1}). This is true when the background velocity model is identical between the compared waveforms, namely with and without the topography model. In this case, we use PREM which is a 1-D model and is also consistent with the ray theoretical predictions made for the given topography model. 
\par
Figures $\ref{fig:Pwaves-1D-models-RTWF}$ and $\ref{fig:Swaves-1D-models-RTWF}$ show that the agreement between a ray theoretical prediction of time anomaly caused by topography and the equivalent measurement on full-waveform synthetics is better for \emph{S}-wave phases than for \emph{P}-wave phases. 
More specifically, \emph{ScS, SKS, SKKS} (and \emph{PKP} from the \emph{P}-waves) traveltime shifts due to topography seem to be better predictable with ray theory. However, it should be noted that the difficulty in predicting an adequate time anomaly may not be solely due to ray theoretical limitations. Additional difficulties may also arise when trying to isolate the \emph{P}-waves.
\par
Focusing now on the comparisons of ray theoretical predictions to measurements on full-waveforms using S20RTS as a background model (blue dots on Figures $\ref{fig:Pwaves-1D-models-RTWF}$ and $\ref{fig:Swaves-1D-models-RTWF}$), we see that the agreement between RT and WF is worse than when using PREM as a background model.
Table \ref{tab:dt_regression} summarises the slope values, for example, for the \emph{ScS} phase, deviations from a perfect agreement between RT and WF are $\sim$1.6 to $\sim$3.5 times larger (i.e., 5\% to 18\%) for S20RTS as a background model than for PREM, depending on the topographic model. The slopes vary significantly with the topographic model, and seem to be larger for models with bigger absolute topographic variation or smaller-scale amplitudes of CMB topography. 
\par
\begin{figure}
    \centering
    \includegraphics[width=\textwidth,height=0.5\textheight]{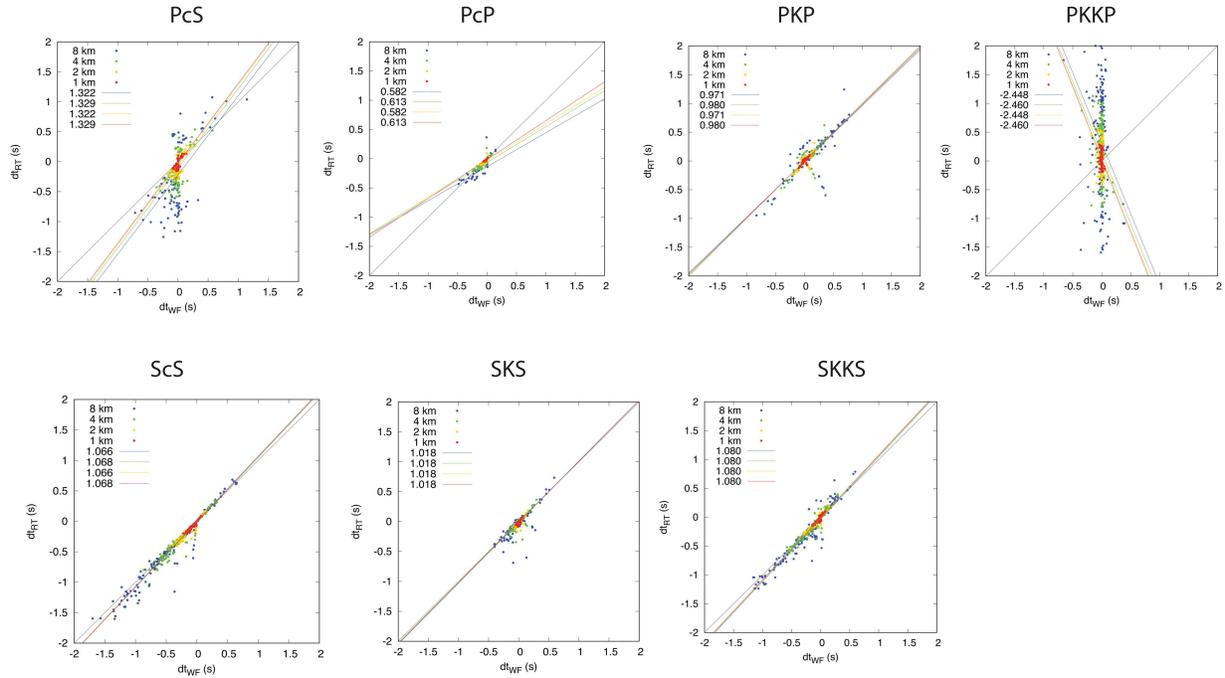}
     \caption{A test using four version of the same topography model to check whether the amplitude of the variation linearly affects the RT prediction and cross-correlation time shift measurements. The model \emph{{LM91}} is scaled for $\pm$1 km (yellow dots), $\pm$2 km (orange dots), $\pm$4 km (green dots) and $\pm$8 km (blue dots) topographic variations. The slopes of a simple linear regression for the cloud of points remain almost unchanged with the scaling of the model. This indicates that there is no clear dependence on the topographic variation when considering the agreement between prediction of RT and CC measurement.}
    \label{fig:scaledli}
\end{figure}

We further investigated this behaviour by computing traveltime residuals for three scaled versions of model \emph{LM91}, with maximum topographic amplitudes of $\pm$1~km, $\pm$2~km, and $\pm$4~km, in addition to the original \emph{LM91} model, whose maximum amplitude is $\pm$8~km. These are shown in Figure $\ref{fig:scaledli}$.
The results for the case of a 3-D background model show that the slope values are very close for all the scaled versions and the original model. 
This implies that the amplitude of the topography has some effect, but that it does not crucially affect the agreement between RT and WF, at least for amplitudes up to $\pm$8~km. The differences in slopes for different topographic models could therefore be due to differences in the scale of the topographic anomalies. 
We note that S20RTS is a relatively smooth 3-D model compared to more recent global tomographic models. We expect even larger discrepancies for 3-D models with stronger, smaller-scale velocity anomalies (e.g., S40RTS \citep{s40rts} or SEMUCB-WM1 \citep{French2014}).
\par
Our results illustrate the key role played by 3-D velocity variations in determining traveltimes. While ray theoretical predictions are based on a 1-D ray tracing expression, the result remains that the measurements on full-waveform synthetics show discrepancies with predictions, even though the measurement is done by comparing 3D and 3D+CMBtopo synthetics (keeping a fixed velocity model), indicating a non-linear effect of the two parameters to the traveltime. 
This also shows the highly non-linear effects of topography and 3-D variations on traveltimes of these phases, when topography is the only parameter added to the simulation. It is expected that some of the mismatch could properly be accounted for when a 3-D velocity correction is made, but it is not expected that it will fully be compensated, due to non-linear trade-off between CMB topography and the 3-D structure of the lowermost mantle, and to finite-frequency effects discussed in the next section.

\begin{table}
    \centering
    \begin{tabular}{l| l| l| l| l}
        \hline
        Phase & T1 & TC1 & TK10 & LM91\\
        \hline
        phase & T1 & TC1 & TK10 & LM91\\
        ScS & 1.18/1.07 & 1.08/1.05 & 1.05/1.02 & 1.07/1.02\\
        SKS & 0.88/0.79 & 1.09/0.79 & 0.26/0.87 & 1.02/0.97\\
        SKKS & 1.14/1.16 & 1.03/1.17 & 0.96/1.19 & 1.08/1.11\\
        PcP & 0.90/1.08 & 1.05/1.11 & 1.06/1.14 & 0.97/1.19\\
        PKP & 1.09/1.16 & 0.80/0.89 & 0.89/1.20 & 0.97/1.00\\
        PKKP & 0.34/0.42 & 0.07/-0.64 & -0.83/-1.27 & -2.45/-0.91\\
        PcS & 2.13/1.47 & 1.21/1.30 & 1.39/1.30 & 1.32/1.29\\
        \hline
    \end{tabular}
    \caption{Slopes of the linear regression lines for the $dt_{RT}$ vs. $dt_{WF}$ diagrams in Figures. \ref{fig:Pwaves-1D-models-RTWF} and \ref{fig:Swaves-1D-models-RTWF}. Each cell shows two slopes, which correspond to S20RTS (left side of the forward slash), and PREM (right side) as a background model, respectively.}
    \label{tab:dt_regression}
\end{table}

\section{FR\'ECHET KERNEL GALLERY FOR CORE-MANTLE BOUNDARY RELATED PHASES}

\begin{figure}
    \centering
    \includegraphics[width=0.95\textwidth, height=0.75\textheight]{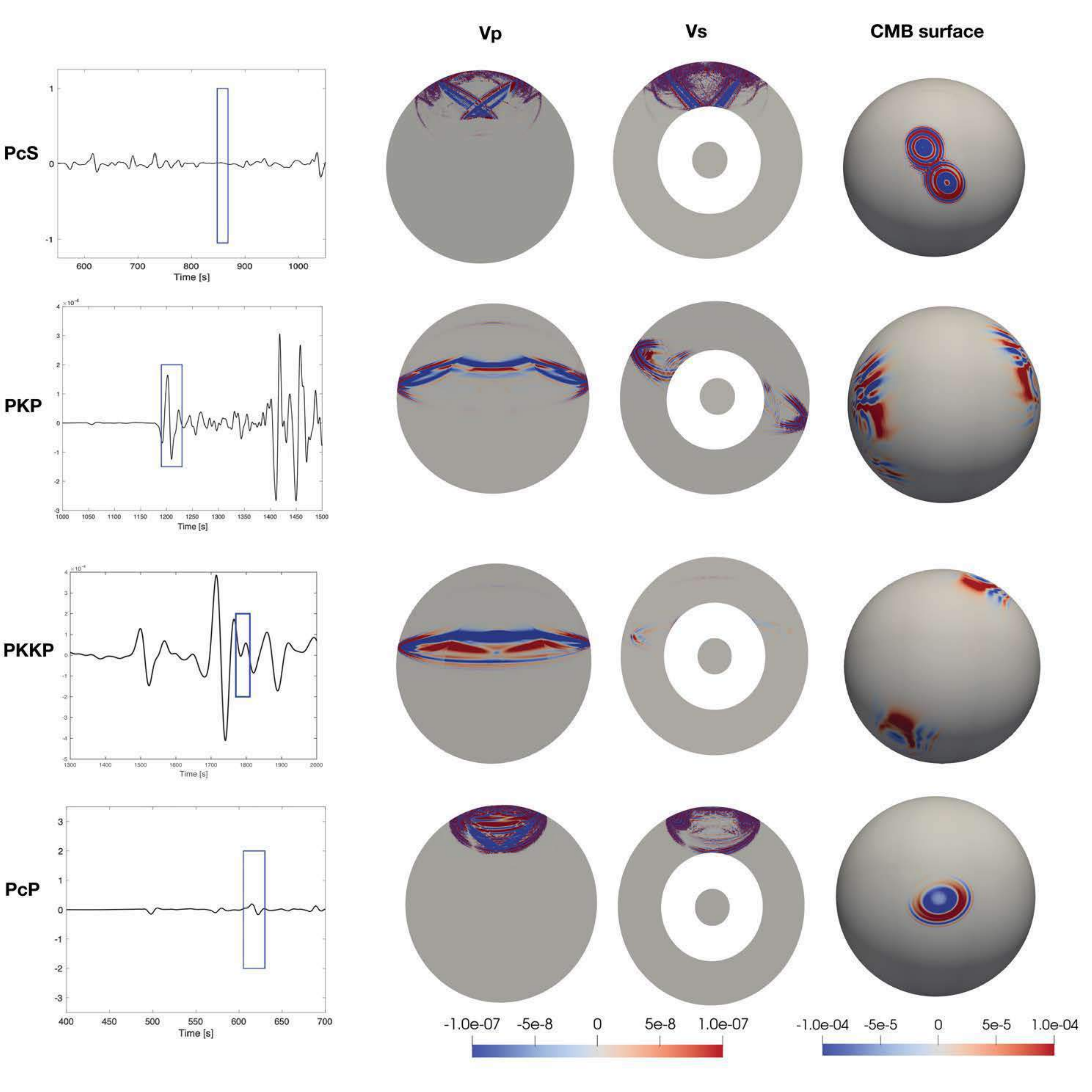}
    \caption{Fr\'echet sensitivity kernels for the group of \emph{P-}waves examined here. Each row corresponds to a phase, from left to right, the figures show: synthetic seismogram and corresponding time window isolating the phase, compressional wavespeed volumetric kernel ($V_{P}$), shear wavespeed volumetric kernel ($V_{S}$), boundary kernel on denoting the core-mantle transition. From top to bottom: \emph{PcS (53$^{\circ}$), PKP (160$^{\circ}$), PKKP (110$^{\circ}$), PcP (53$^{\circ}$)}. The units of volumetric kernels are $km^{-3} \cdot s$ and for the boundary kernels $km^{-2} \cdot s$.}
    \label{fig:pkernels}
\end{figure}

Figure $\ref{fig:pkernels}$ shows the complete sensitivity of the \emph{P-}wave phases under investigation.
The Fr\'echet derivatives are computed for compressional ($V_{P}$) and shear wavespeed ($V_{S}$) as well as for boundary of the interface corresponding to core-mantle boundary at approximately 2890 km for the chosen background model, PREM.
All \emph{P}-wave phases show their theoretical, finite-frequency, hence broader Fresnel zone, sensitivity to $V_{P}$. This indicates that the time window chosen around the predicted arrival according to PREM encapsulates the phases well enough and they present little or no interference with other waves.
It is worth noting that, specifically for the \emph{PcS} and \emph{PcP} phases, they seem to exhibit considerable sensitivity to shear wavespeed as well, showing the more complex contributions to their traveltimes and perhaps explaining the weakening of their sensitivity to topography, as shown in the analysis done in section $\ref{sec:dts}$. 
\par
\begin{figure}
    \centering
    \includegraphics[width=\textwidth,height=0.65\textheight]{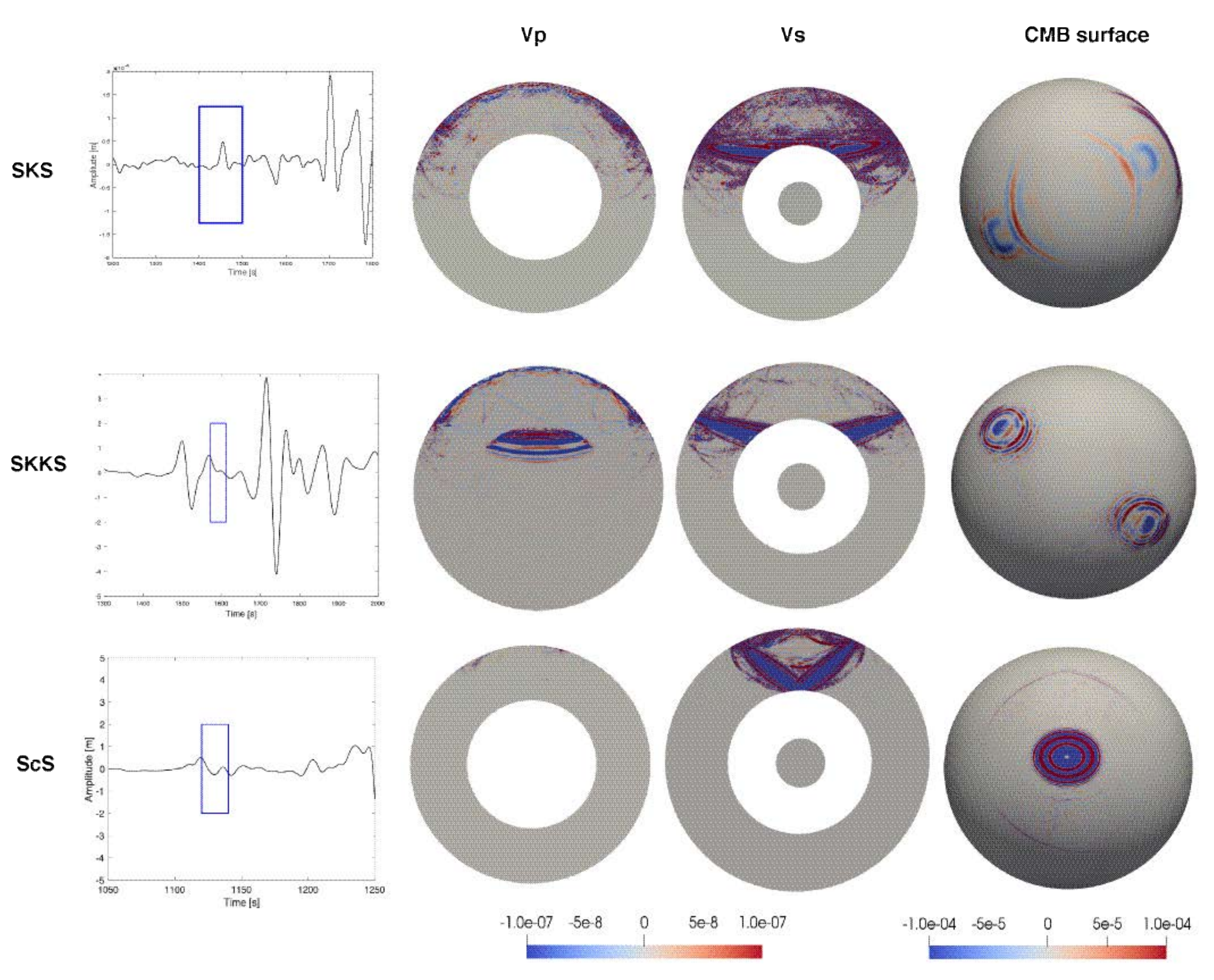}
     \caption{Similar to figure $\ref{fig:pkernels}$, now for the selected \emph{S-}wave phases. From top to bottom: \emph{SKS (110$^{\circ}$), SKKS (110$^{\circ}$) and ScS (53$^{\circ}$)}. The units of volumetric kernels are $km^{-3} \cdot s$ and for the boundary kernels $km^{-2} \cdot s$.}
    \label{fig:skernels}
\end{figure}

Regarding the \emph{S-}wave phases, displayed in Figure $\ref{fig:skernels}$, their traveltime sensitivity to shear and compressional wavespeeds and CMB surface are shown in a similar way as in Figure $\ref{fig:pkernels}$. It is readily observed that the shear wavespeed kernels computed for the time windows around the predicted arrival of these phases also clearly show their theoretical path, in this case broader and more extended due to the finite frequency effects which are in this case properly accounted for. 
The seismic phase \emph{SKS} shows a broad and noisy sensitivity. The travelling paths of \emph{SKS, SKKS} through the mantle are very similar, with their major differences appearing within the outer core. This is why their differential time has frequently been used for shear wave splitting studies in the lowermost mantle, \citep[e.g.][]{Restivo2006,FaCh03,SiPaTrTr08,SoPo91}. In the context of our study, it is important to understand their finite-frequency sensitivity and its variability between the two phases when it comes to the CMB surface. The \emph{S-}wave phases also show clear contributions only from their theoretically predicted path in the time window selected for the calculation. In contrast to the \emph{P-}wave kernels, there is little to none sensitivity of these phases to compressional wavespeed.
\par
Turning to the boundary kernels, from visual inspection it becomes apparent that contributions to the traveltime sensitivity are low, especially for the traveltime windows of \emph{PKP} and \emph{PKKP}. This is in reference to the background model PREM. A low sensitivity to that means that the theoretical traveltime has low values at the interface of the core-mantle boundary, hence the particular seismic phase does not seem to arrive at a considerably different time than that reference traveltime prediction using PREM, due to the interaction with the core-mantle boundary and its topography.
The boundary sensitivity of \emph{PcS} and \emph{PcP} are more pronounced and seem to sample its structure adequately and with the typical top-side reflection Fresnel zone. Considering the boundary sensitivity of \emph{S-}waves, from the kernels in Figure $\ref{fig:skernels}$, we observe that also for these phases, the reflected \emph{ScS} phase has a clear sensitivity to the CMB. However, for \emph{SKS} and \emph{SKKS}, as shown by their corresponding boundary kernels, most of the sensitivity is observed at the piercing points in and out their way to the core, with \emph{SKKS} having a stronger contribution. We discuss these results in greater detail in the next section.

\section{DISCUSSION \& CONCLUDING REMARKS}

\par
We have presented a traveltime analysis, where a ray theoretical prediction of traveltime anomaly caused by core-mantle boundary topography variations was compared to a measurement made by cross-correlation on full-waveform synthetics, imitating the common techniques of measuring time shifts on recorded waveforms to infer CMB topography. The retrieval of traveltime anomalies using ray theory shows that we could predict the shift caused by a given topography variation, albeit with some loss of accuracy and most reliably in time windows for the selected \emph{S-}wave phases. We also showed that the correspondence worsens when the measurements are made in a 3-D background model, implying considerable effects of lateral velocity variations on the traveltimes. This is shown by a deviation from a perfect linear correspondence between the prediction and the measurement on realistic waveforms.
\par
For the \emph{P-}wave phases, the correspondence between ray theory and measurements on waveforms is much worse than for \emph{S-}wave phases, except for \emph{PKP} and \emph{PcP}. The lack of agreement could be associated with better time window isolation of the given \emph{S-}wave phases, contrary to the difficulties occurring for \emph{P} waves. Specifically, multiple branches of \emph{PKKP} can make its peak separation from its different branches (e.g. \emph{PKKPdf}) more difficult than for other phases. It is also generally expected that \emph{P}-wave propagation is more complex when interaction with strong velocity contrasts occurs \citep{AkRi80,DoMo80}. This involves both up- and down- going \emph{P} and \emph{S} waves when a \emph{P}-wave interacts with an interface, and especially the CMB, due to the strong contrasts in density and thermo-elastic properties between the solid rocks defining the mantle and the liquid iron alloy determining outer core composition.
\par
\begin{figure}
    \centering
    \includegraphics[width=\textwidth, height=0.65\textheight]{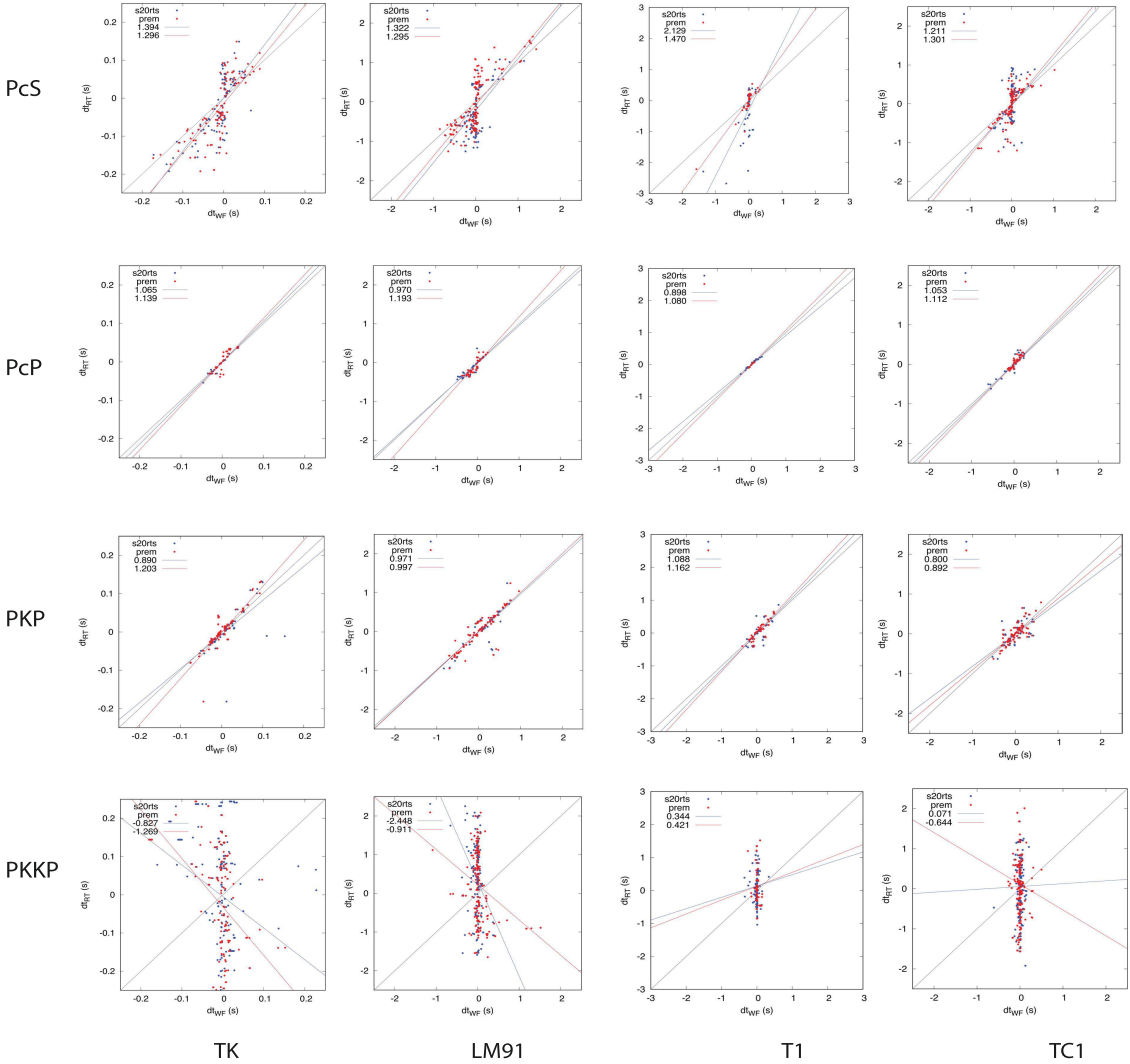}
    \caption{Comparison between ray theoretical prediction of time shift given a CMB topography model to the time anomaly measured on each time windowed seismic phase by comparing synthetic waveforms in \textcolor{red}{1-D} and \textcolor{blue}{3-D} velocity background, red and blue points on the plots, respectively. Each column corresponds to a different topography model (\emph{{TK, LM91, T1, TC1}}). Each row corresponds to a seismic phase (\emph{PcS, PcP, PKP} and \emph{PKKP}). Most of the phases show acceptable correspondence between predicted and measured by cross-correlation time shift, implying that the signature of CMB topography can be only partially captured by using RT. This is not the case for phase \emph{PKKP}, where a prediction fails completely and the time window isolation is insufficient to make a reliable measurement.}
    \label{fig:Pwaves-1D-models-RTWF}
\end{figure}
\begin{figure}
    \centering
    \includegraphics[width=\textwidth, height=0.5\textheight]{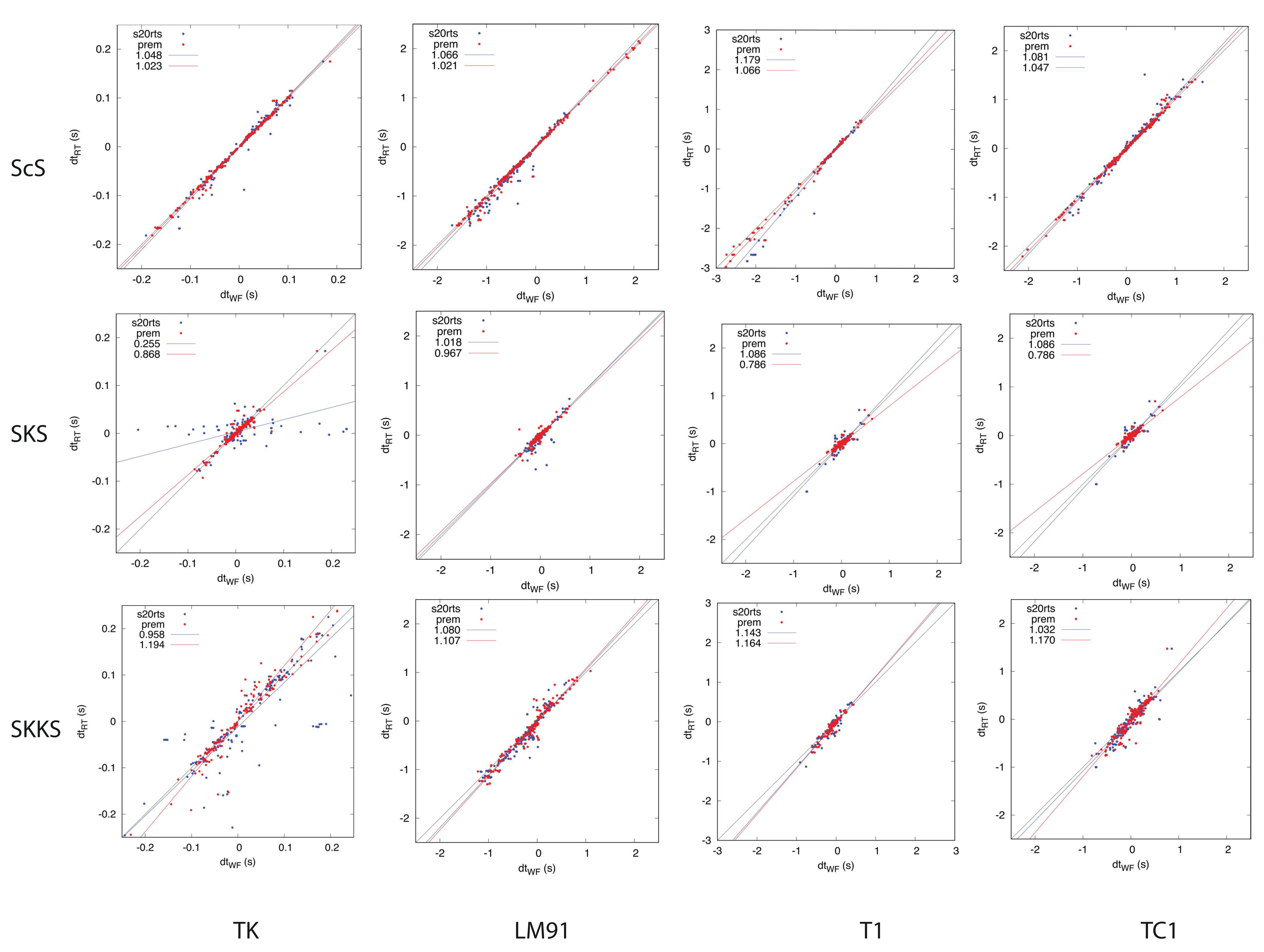}
    \caption{Similar as figure $\ref{fig:Pwaves-1D-models-RTWF}$. Each row corresponds to a seismic phase (\emph{ScS, SKS} and \emph{SKKS}). Most of the phases show better correspondence between predicted and measured by cross-correlation time shift, implying that S-wave phases have a better sensitivity to CMB topography and this can be accommodated by RT.}
    \label{fig:Swaves-1D-models-RTWF}
\end{figure}

For both the case of PREM (Figures $\ref{fig:Pwaves-1D-models-RTWF}$ and $\ref{fig:Swaves-1D-models-RTWF}$ red points) and S20RTS (Figures $\ref{fig:Pwaves-1D-models-RTWF}$ and $\ref{fig:Swaves-1D-models-RTWF}$ blue points) as a background model, we observe deviations from a perfect agreement between traveltime residuals predicted by ray theory, and those measured using cross-correlation on realistic waveforms (slope value equal to unity). 
The deviations seem to increase linearly with the amplitude of the CMB topography. The rate of increase does not seem to depend on the amplitude of the CMB topography, but varies with topographic models, which could indicate a dependence on the scale of topographic anomalies. In particular, the rate seems larger for CMB topography models \emph{T1} and \emph{TC1} \citep{Deschamps2018}, which have smaller-scale anomalies, but are not built on seismic data and whose geographical distribution may thus differ significantly from the real CMB topography. The deviations for the case of PREM as a background model are between 2\% to 7\% for the \emph{ScS} phase (the phase with the most reliable measurements), and between 5\% and 18\% for the case of S20RTS as a background model. 
\par
The slopes given in Table $\ref{tab:dt_regression}$ for each of the phases and each of the background model show that the sign of the topographic variation can be in most cases correctly defined. However, for the \emph{PKKP} phase and all models tested in our study, the measurement done in both background models seems to fail to infer the correct sign of topography (negative slopes). The linear regression attempted to fit the points of measurement, though, is not efficient as there are many points for which the ray theoretical prediction is close to zero. This perhaps implies that \emph{PKKP} is indeed difficult to isolate and given its multiple branches, an assignment to its traveltime shift due to CMB topography, neglecting effects from lowermost mantle and outer core velocity variations, is a rather simplified assumption as many effects can lead to almost synchronous arrivals \citep[e.g.][]{EaSh97,RoRe03}. We note that difficulties isolating separate branches of the \emph{PKKP} phase would be reduced when using higher-frequency waveforms, as usually done when using ray theory, \citep[e.g.][]{Tanaka2010}. This is, however, not currently possible when using our spectral-element synthetics, because of the significant increase in computational requirements for higher frequencies. This implies that phases such as high-frequency \emph{PKKP} are currently difficult to use in full-waveform inversion for the core-mantle boundary topography.
\par
The smaller deviations when using PREM as a background model suggest that the ray theory expressions in eqs. \ref{eq:rt_kernel_ref} and \ref{eq:rt_kernel_tra} are not completely adequate for the time shift measurements on 3-D background. These are derived using the ray parameter for a 1-D model (i.e., without topography), which implies that the observed deviations could be an effect of the small change in ray parameter due to the addition of CMB topography. Most likely though, the deviation could be due to finite-frequency effects when using the full-waveform theory.
The larger deviations observed when using S20RTS as a background model indicate trade-offs between the CMB topography, and 3-D \emph{S-} and \emph{P-} velocity anomalies in the mantle. For model \emph{T1} with strong, small-scale topographic anomalies beneath cold downwellings at the CMB, the deviations reach 18\% for the \emph{ScS} phase. This could affect the accuracy of inferred global tomographic models, which often do not account for the core-mantle boundary topography, although deviations in our calculations are not so large as to imply significant artifacts in tomographic models. 
\par
The finite-frequency sensitivity kernels show that the phases we investigated have strong contribution in their predicted traveltime windows and are not interfering with other phases. This observation shows that measurements on these phases can contribute to a good quality traveltime dataset and misfit measurements on their time windows and could help for a meaningful model update, given that they can be well separated with time windows. 
We also observe that reflected \emph{P}-wave phases, i.e. \emph{PcP} and \emph{PcS}, are sensitive to shear wavespeed structure, too. In common inversion methods for imaging CMB topography, traveltimes are corrected for mantle velocity structure. The traveltime sensitivity of the \emph{S}-waves does not seem to present similar patterns to that of \emph{P}-waves, namely the traveltime is affected almost only from shear wavespeed variations. This may partly explain the poorer agreement between ray theory and measured traveltimes for \emph{P-}waves than for \emph{S}-waves, as the ray theoretical prediction does not account for the influence of shear wavespeed structure on the \emph{P}-wave traveltimes (for example see expression $\ref{eq:rt_kernel_ref}$). It also implies that linearised corrections for mantle structure may require the correction for shear wavespeed structure as well, similar to a conclusion made by \citep{KoPaTr19} for the, rather difficult to observe, \emph{PP} precursors.
\par
The sensitivity of the traveltimes to CMB seems to be weak for most of the selected time windows, implying that the contribution of the CMB interface to the traveltime shift is too weak to provide a robust update of the topography. Notable exceptions are the phases which reflect off the CMB, namely the \emph{ScS, PcP, PcS} and to a lesser extent \emph{SKS, SKKS} at the piercing points on the way in and out the outer core. 
This indicates that for purely CMB topography retrieval using traveltimes of these phases, one should consider that their sensitivity differs significantly and some phases offer a more optimal illumination of CMB structure than others (for example compare \emph{ScS} to \emph{SKS}). 
\par
\begin{figure}
    \centering
    \includegraphics[width=0.95\textwidth,height=0.65\textheight]{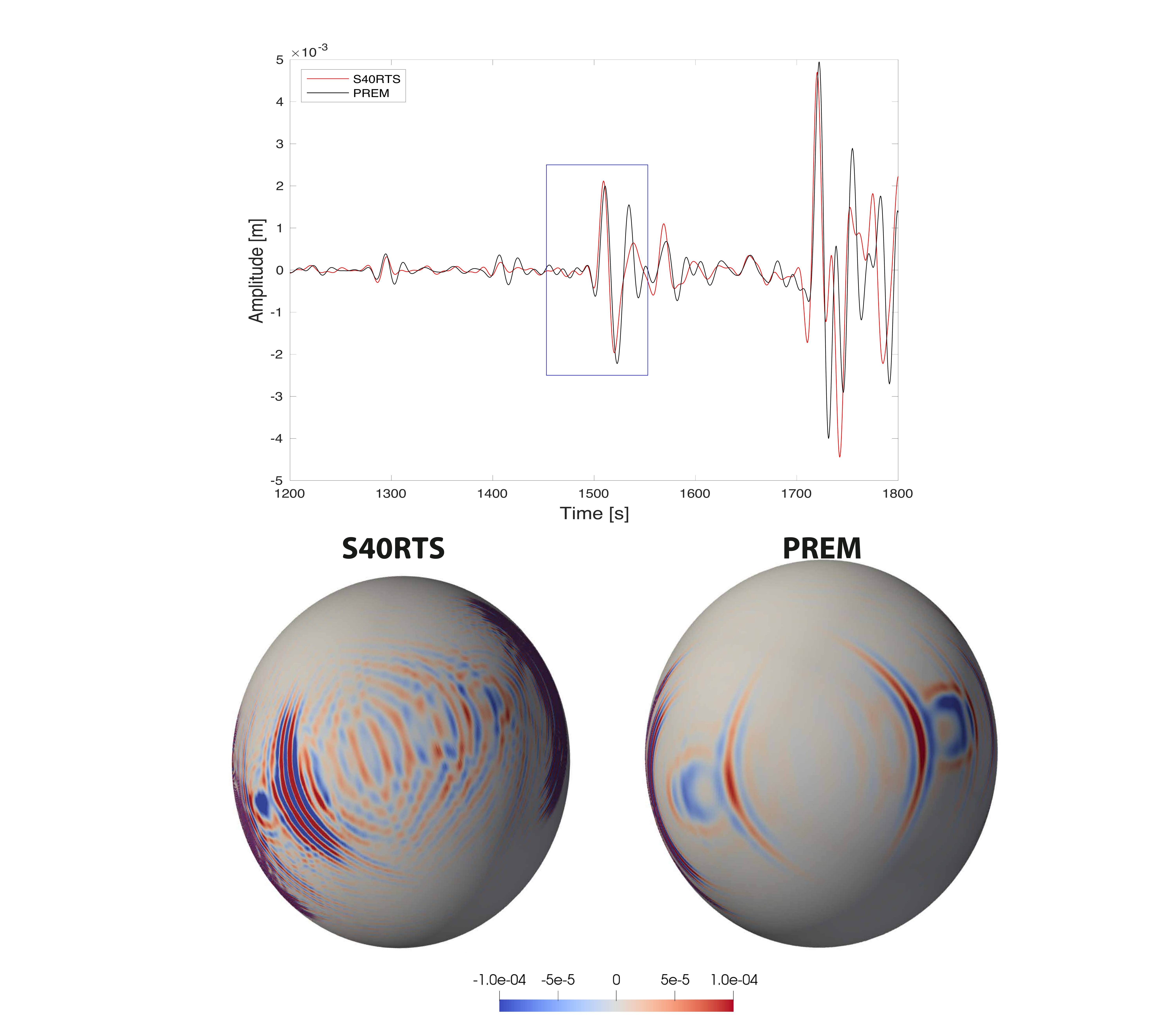}
     \caption{Comparison between the boundary kernels computed in 3-D background using model S40RTS (red line) and the same seismic phase \emph{SKS} in 1-D background PREM (black line). the fact that the boundary kernels differ due to the different background model shows that interpretation of the traveltime shift of the phases due to topography should take into account the trade-off between topography and 3-D velocity variation, ideally in a non-linear way (not only by subtracting a time shift correction due to the 3-D mantle model). The units of boundary kernels $km^{-2} \cdot s$.}
    \label{fig:1dvs3dSKS}
\end{figure}

Trying to explain the discrepancies between predicted and measured traveltime anomaly when working in 3-D background, this time in PREM and S40RTS \citep{s40rts} (1-D and 3-D mantle models), we computed boundary sensitivity kernels for the same time window isolating the phase \emph{SKS}, same station-source configuration as in Figure $\ref{fig:Swaves-1D-models-RTWF}$. This phase is chosen for exemplification. The resulting kernels show that the 3-D model in the background significantly affects the boundary sensitivity of the traveltime, see Figure $\ref{fig:1dvs3dSKS}$.
This indicates that 3-D effects are imprinted into the traveltime of seismic phases and their sensitivity to the CMB is clearly changing in a substantial way, thus making an interpretation of the traveltime anomaly due to only topographic variations difficult. Additionally, given that our knowledge for the lowermost mantle structure is still lacking sufficient resolution, a joint inversion using Fr\'echet kernels, where both velocity and CMB topography are updated, may be more appropriate for the nature of this problem.
This also confirms the traveltime analysis comparing traveltimes calculated using ray theory to those measured on synthetics, and indicates that not addressing this trade-off could bias both the resulting velocity structure and the inferred CMB topography. 
\par
From our sensitivity analysis we can conclude that top reflected seismic phases (e.g. \emph{ScS}) are adequately informative for properly imaging CMB topography using boundary kernels. We also find that the effect of the velocity variations are significant and should be taken into account using the volumetric finite-frequency kernels as shown in this work. To reliably separate the two effects seems complicated, especially with an one step approach; however, it can be more efficient to utilise an iterative optimisation of a starting model and help make better inferences about CMB topography. High resolution imaging of the CMB and the mantle structure just above this interface requires an interdisciplinary approach. Understanding how subduction processes and hot material accumulated at the bottom of the mantle can affect observations of seismic phases traveltimes on the recordings is fundamental for a clarified selection of seismic traveltime data with stronger sensitivity at these regions.
\clearpage


\clearpage

\clearpage

\begin{figure}
    \centering
    \includegraphics[width=\textwidth]{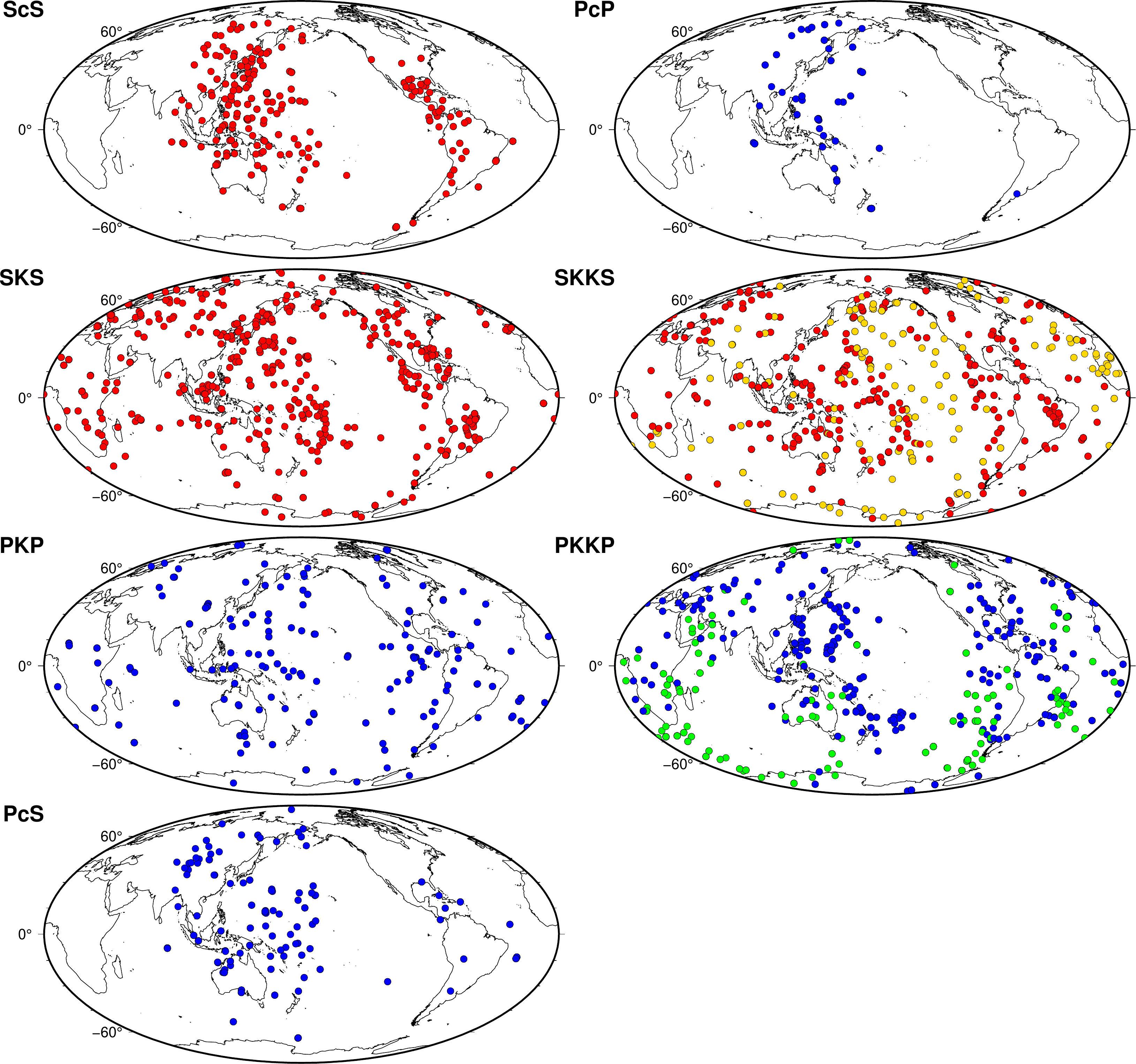}
    \caption{Bouncing or crossing points at the CMB for the seven seismic phases used in this study, as labeled in each panel. Blue-filled circles show P phases, while red circles show S phases. For SKKS and PKKP, yellow- and green-filled circles show the underneath bouncing points at the CMB.}
    \label{fig:bouncing}
\end{figure}
\clearpage

\clearpage

\clearpage

\clearpage

\clearpage

\clearpage

\clearpage

\clearpage


\bibliographystyle{unsrtnat}
\bibliography{references} 

\end{document}